\documentclass[preprint2]{aastex}

\newcommand{\myemail}{shogo@optik.mtk.nao.ac.jp}

\newcommand{\uprule}{\rule{0pt}{3.5ex}}
\newcommand{\dorule}{\rule[-2ex]{0pt}{3.0ex}}

\shorttitle{Extinction Law $VIJHK_S$}
\shortauthors{Nishiyama et al.}

\begin{document}

\title{Interstellar Extinction Law toward the Galactic Center I\hspace{-.1em}I:\\
$V$, $J$, $H$, and $K_S$ Bands}

\author{Shogo Nishiyama\altaffilmark{1}, 
Tetsuya Nagata\altaffilmark{2}, 
Motohide Tamura\altaffilmark{1}, 
Ryo Kandori\altaffilmark{1},
Hirofumi Hatano\altaffilmark{3},
Shuji Sato\altaffilmark{3},
and Koji Sugitani\altaffilmark{4}
}

\altaffiltext{1}{National Astronomical Observatory of Japan, 
Mitaka, Tokyo 181-8588, Japan; \myemail}

\altaffiltext{2}{Department of Astronomy, Kyoto University, 
Kyoto 606-8502, Japan}

\altaffiltext{3}{Department of Astrophysics, Nagoya University, 
Nagoya 464-8602, Japan}

\altaffiltext{4}{Graduate School of Natural Sciences, Nagoya City University,
Nagoya 467-8501, Japan}

%------------------------------------------------------

\begin{abstract}

We have determined the ratios of total to selective extinction directly from observations
in the optical $V$ band and near-infrared $J$ band toward the Galactic center.
The OGLE (Optical Gravitational Lensing Experiment) Galactic bulge fields
have been observed 
with the SIRIUS camera on the IRSF telescope, and 
we obtain $A_V/E_{V-J}=1.251\pm0.014$ and $A_J/E_{V-J}=0.225\pm0.007$. 
From these ratios, we have derived $A_{J}/A_{V} = 0.188 \pm 0.005$; 
if we combine $A_{J}/A_{V}$ with the near-infrared extinction ratios
obtained by Nishiyama et al. for more reddened fields near 
the Galactic center,
we get
$ A_{V} : A_{J} : A_{H} : A_{K_S} = 1 : 0.188 : 0.108 : 0.062$, 
which implies steeply declining extinction toward the longer wavelengths.  
In particular, it is striking that 
the $K_S$ band extinction is $\approx 1/16$ of the visual extinction $A_V$,
much smaller than one tenth of $A_V$ so far employed.

\end{abstract}

\keywords{dust, extinction --- stars: horizontal-branch --- Galaxy: center}

%------------------------------------------------------

\section{INTRODUCTION}
\label{sec:Intro}

The wavelength dependence of interstellar extinction provides 
important diagnostic information about the dust grain properties.  
Interstellar extinction law shows a large range of variability 
from one line of sight to another, especially in the ultraviolet 
and optical wavelengths.  
In comparing the wavelength dependence among different lines of sight, 
the normalization of the extinction curves by the total extinction 
$A_V$, instead of the usual color excess $E_{B-V}$, is vitally important, 
as \citet[][CCM]{CCM89} have shown.  
According to CCM, 
the variation in Galactic extinction curves from the ultraviolet to 
the optical is described by a single parameter, 
which itself is the ratio of total to selective extinction 
$R_V = A_V / E_{B-V}$.  
Thus the ratio $R_V$, or more generally  
$R_{\lambda} = A_{\lambda}/E_{{\lambda}'-{\lambda}}$, 
is very important, although very difficult to obtain.  
An usual way to determine $R_V$
is to extrapolate the ratio of color excesses 
$E_{\lambda-V}/E_{B-V}$ to $\lambda = \infty$ 
with reference to {\it some model}, 
and can be compromised by {\it emission} or {\it scattering} 
by dust grains near the stars.  

The past decade has seen a 
new method to determine the ratio of total to selective extinction 
$R_{\lambda}$.  
The method was first 
proposed by \citet{Woz96}; 
in essence, it simply measures the regression of the mean color of
red clump (RC) stars in the Galactic bulge on their mean magnitude
(RC method).  
The increase in the amount of dust in a line of sight causes 
the clump fainter and redder.  
The slope of these changes in 
a color-magnitude diagram (CMD) is equivalent to $R_{\lambda}$.  
The method has been developed by \citet{Stanek96}, \citet{Udal03}, 
and \citet{Sumi04} in the $V$ and $I$ bands\footnote{They regard their 
$I$ filter as similar to the Landolt Cousins $I$, whose effective 
wavelength is about $0.80\mu$m.  
}, 
and recently applied to the near-infrared wavebands 
by \citet[][hereafter paper I]{Nishi06}.

As a result, 
a somewhat surprising suggestion was first made by \citet{Popo00} 
for the interstellar extinction toward the Galactic bulge;
the ratio of total to selective extinction $R_{VI} = A_V / E_{V-I}$ is 
approximately 2.0, 
much smaller than the ratio 2.5 for the standard ($R_V = 3.1$) CCM 
extinction curve \citep{Udal03}.  
This was confirmed by \citet{Sumi04} in a larger number of fields.  
Thus the $V$ ($0.55 \mu$m) to $I$ ($0.80 \mu$m) part 
of the extinction curve toward the Galactic bulge 
seems to be characterized by a smaller $R_{VI}$.  %%(``steep curve'').
In the near infrared, the extinction law is frequently referred to 
as ``universal'', and in fact CCM gives an $R_V$-independent curve.  
However, paper I have shown that the near-infrared extinction curve 
toward the Galactic Center (GC) is different from those in the literature, 
having also smaller 
$A_{H} / E_{J-H}$ and $A_{K_S} / E_{H-K_S}$.

In this paper, we apply the RC method and extend the $R_{\lambda}$ 
determination to the $J$ band ($1.25 \mu$m) 
by measuring the $J$ magnitude of the stars whose $V$ photometry 
was obtained in \citet{Udal03}.  
Furthermore, assuming that the extinction curve can be extended to 
more heavily reddened regions observed in paper I, 
we estimate $A_{\lambda}/A_V$ at $J$, $H$, and $K_S$ wavelengths.

\section{Observation, Data Reduction, and Analysis}
\label{sec:Obs}

\subsection{Near-infrared Observations}

All observations in the near-infrared wavelengths were made with the SIRIUS 
camera 
\citep[Simultaneous InfraRed Imager for Unbiased Survey;][]{Nagas99, Nagay02}  
attached to the 1.4 m telescope IRSF (InfraRed Survey Facility),
on the nights of 2004 May 18 and 19.
The SIRIUS camera
provides photometric images 
of a 7\farcm7 $\times$ 7\farcm7 area of sky 
in three near-infrared wavebands $J$ ($1.25\mu$m), $H$ ($1.63\mu$m),
and $K_S$ ($2.14\mu$m) simultaneously.
The detectors are three 1024 $\times$ 1024 HgCdTe arrays,
with a scale of 0\farcs45 pixel$^{-1}$.  
The IRSF/SIRIUS system is similar to the MKO system \citep{Tok02}.

\begin{figure}[t]
 \begin{center}
    \plotone{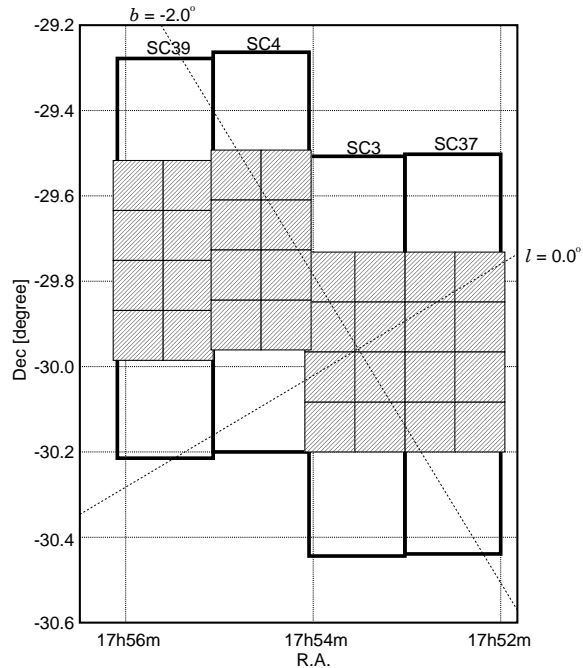}
  \caption{
	Observed area and fields used for data analysis.
	Thick-line rectangles, labeled as ``SC39'', ``SC4'', 
	``SC3'', and ``SC37'' 
	show the areas observed by the OGLE project.
	The fields observed by IRSF/SIRIUS are shown by hatched squares.
	}
  \label{fig:survey}
 \end{center}
\end{figure}

Over the range of $17^{\mathrm{h}}52^{\mathrm{m}} \la \mathrm{R.A.(J2000.0)} \la 
17^{\mathrm{h}}56^{\mathrm{m}}$
and $-30\fdg2 \la \mathrm{Dec.(J2000.0)} \la -29\fdg5$, 
32 images in each band were obtained (Fig. \ref{fig:survey}).  
These fields correspond to the region ``A'' 
($l \approx 0 \degr, b \approx -2 \degr$) of \citet{Udal03} in the 
Optical Gravitational Lensing Experiment (OGLE) I\hspace{-.1em}I maps 
of the Galactic bulge 
in the $V$ and $I$ bands\footnote{
 In this paper, we do not use the $I$ band data,
 because the OGLE $I$ band filter has 
 a wider long-wavelength wing of transmission contrary to
 the sharper drop of the standard $I$ band filter, 
 and the relation between $V$ and $I$ extinction
 has been already studied by \citet{Udal03}, and \citet{Sumi04}.
} 
with an absolute photometric accuracy of 0.01-0.02 mag \citep{Udal02}. 
The data files containing the photometry 
were downloaded from the OGLE website\footnote{
http://bulge.princeton.edu/\texttt{\symbol{`~}}ogle/ogle2/bulge\texttt{\symbol{`_}}maps.html}.  
The \citet{Udal03} ``A'' region consists of 
the five fields SC3, SC4, SC5, SC37, and SC39 
(Fig. \ref{fig:survey}), 
but the SC5 data were excluded from the following analysis 
because it suffers from the heaviest extinction and 
its $V$ magnitudes become unreliable, 
as \citet{Udal03} also pointed out.  

%%%%---------------------------------
%\newpage

The observing weather was photometric, with seeing of $\sim$1\farcs1
in the $J$ band.
Flat fields were obtained during 
each clear evening and morning twilight.
Dark frames were taken at the end of each observing night.
A single image comprises 10 dithered 5 sec exposures.  

The SIRIUS camera provides three ($J$, $H$, and $K_S$) images 
simultaneously, 
and thus we have the data sets of the same region in the three bands.  
However, the uncertainties in the $H$ and $K_S$ photometry are larger, 
and the range of extinction is smaller, about a half ($H$) 
and a third ($K_S$) of that in the $J$ band\footnote{
In paper I,
we have observed fields closer to the GC
where the range of extinction is large (about 3.5 mag in the $J$ band),
and thus we were able to determine 
the extinction ratio in the $J,H$, and $K_S$ bands accurately.
However, the fields observed in this study are located at $b \sim -2\degr$,
and the range of extinction is 0.5 mag in the $J$ band
(see the bottom panel in Fig. \ref{fig:Slopes}),
which is less than 0.2 mag in the $K_S$ band.
}.  
Therefore we have obtained reliable results only in the $J$ band.

%%%%---------------------------------
\begin{figure}[h]
 \begin{center}
  \rotatebox{-90}{
  \epsscale{.60}
    \plotone{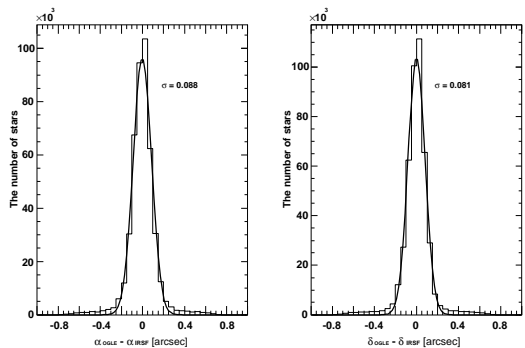}
  }
  \caption{
    Histograms of the positional difference
    between the OGLE and IRSF coordinates
    for cross-identified stars
    in R.A. ({\it left}) and Dec. ({\it right}).
    The values of $\sigma$ obtained by fitting with a Gauss function
    are 0\farcs088 in R.A. and 0\farcs081 in Dec., respectively.
    The positional offsets between 
    the OGLE and IRSF coordinates were already corrected.
  }
  \label{fig:PosDiff}
 \end{center}
\end{figure}

IRAF (Image Reduction and Analysis Facility)\footnote{
IRAF is distributed by the National Optical Astronomy
Observatory, which is operated by the Association of Universities for
Research in Astronomy, Inc., under cooperative agreement with
the National Science Foundation.}
software package was used
to perform dark and flat-field corrections,
followed by sky background estimation and subtraction.
Photometry, including point spread function (PSF) fitting, was carried out 
with the DAOPHOT package \citep{Stetson87}.
We used the DAOFIND task to identify point sources,
and the sources were then input 
for PSF-fitting photometry to the ALLSTAR task.
About 20 sources were used to construct the PSF for each image.

Each image was calibrated with the standard star
\#9172 \citep{Persson98},
which was observed every half an hour.
We assumed that \#9172 is $J=12.48$ 
in the IRSF/SIRIUS system.
The average of the zero point uncertainties
and the 10$\sigma$ limiting magnitude in the $J$ band
were $\sim 0.02$ mag and $16.8$ mag, respectively.

Astrometric calibration was performed, field by field,
with reference to the positions of stars 
in the 2MASS point source catalog \citep{Skrutskie06}.
Only the stars with the photometric error of less than 0.05 mag
in the 2MASS and our catalog were used for the calibration.
The positional difference {was finally calculated
by using the stars of $\la 0.1$ mag photometric error, 
and we obtained
the standard deviation of the positional difference better than 0\farcs1.

\subsection{Cross-identification of the SIRIUS and the OGLE source}

The stars found with IRSF/SIRIUS and OGLE 
have been cross-identified, field by field, 
using a simple positional correlation.
The procedure consists of two steps.
First, identification was performed with large radius (1\farcs5)
to evaluate astrometric offset
between IRSF and OGLE coordinates.
We found that the offsets 
in R.A. and Dec. are 
in the range from $-0\farcs7$ to $+0\farcs1$, 
and from $+0\farcs1$ to $+0\farcs4$, 
respectively.
The offsets seem to depend on the position in the OGLE field.
Second, the OGLE coordinates were corrected for these offsets,
and a search radius of 0\farcs7 was used for the identification.

We show the histograms of positional differences 
of the finally identified stars in Fig. \ref{fig:PosDiff}.
We found $\sim 25,000$ matches in each field with an rms error 
in the difference between SIRIUS and OGLE coordinates
0\farcs13 in R.A. and Dec.
The values of $\sigma$ obtained by fitting with a Gauss function
are 0\farcs088 in R.A. and 0\farcs081 in Dec.

\subsection{Data Analysis}

To measure the ratios of total to selective extinction
$A_V/E_{V-J}$ and $A_J/E_{V-J}$,
we selected the bulge RC stars,
which constitute a compact and well-defined clump in a CMD, 
and are thus good tracer of extinction and reddening.
The $V$ versus $V-J$ CMD constructed with the IRSF and OGLE data
is shown in Fig. \ref{fig:CMD}.
In this analysis we follow the procedure described
in \citet{Udal03}, \citet{Sumi04}, and paper I.

%%%%---------------------------------
\begin{figure}[h]
 \begin{center}
  \rotatebox{-90}{
  \epsscale{0.9}
    \plotone{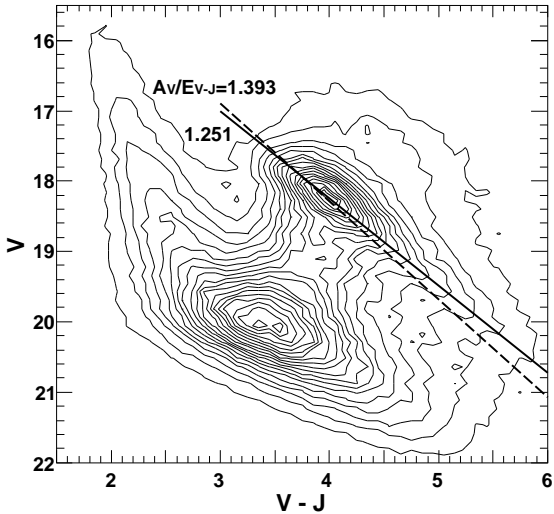}
  }
  \caption{
    $V$ versus $V-J$ color magnitude diagram of the OGLE bulge fields.
    Contours are from 0 to 1600, linearly spaced by 80.
    The {\it solid} and {\it dashed} lines show
    the direction of the interstellar extinction
    with $A_V/E_{V-J}=1.251$ (this work),
    and 1.393 \citep[][$R_V=3.1$]{CCM89}, respectively.
  }
  \label{fig:CMD}
 \end{center}
\end{figure}

First, we divide each field of SIRIUS into nine subfields
of $\sim 2\farcm8 \times  2\farcm8$ on the sky.
Then we construct $V$ versus $V-J$ and $J$ versus $V-J$ CMDs for each subfield.
Second, in the CMDs, we extract stars in the region dominated by RC stars,
and the stars are used to make magnitude histograms 
(luminosity function, see also Fig.2 in paper I).
The peaks of RC stars are fitted with a Gaussian function.
Third, the stars in the range fitted above are employed
to see the distribution of RC stars in the color $V-J$, and 
the color peaks of RC stars are also fitted with a Gaussian function.
Since the mean $V$ magnitudes of RC stars become too faint 
in highly reddened fields,
estimates of the peak magnitudes and the colors of RC stars
can be unreliable in such fields.
To avoid this problem, we do not use the subfields
in which we could not find a clear peak of RC stars.  
Note that this exclusion was made in the second and third steps, 
so a field employed in the 
$J$ versus $V-J$ CMD can be excluded 
in the $V$ versus $V-J$ CMD 
if its $V$ distribution does not have a clear Gaussian peak, 
and vice versa.  
As a result, 48 ($V$ versus $V-J$) 
and 8 ($J$ versus $V-J$) out of 288 ($32 \times 9$) subfields 
were excluded from our analysis.
Therefore the data points of the two CMD are not identical, and 
the resultant slopes do not necessarily satisfy the relation 
$A_{V}/E_{V-J} = A_{J}/E_{V-J} + 1$\footnote{ 
We made a $J$ versus $V-J$ CMD excluding 48 subfields,
which are not used in $V$ versus $V-J$,
and confirmed that the exclusion of them 
changes the slope in the CMD
by only 0.002.
}.

\section{Results}
\label{sec:Res}

Combining the IRSF $J$ band and the OGLE $V$ band data sets,
we have made the $V$ versus $V-J$ ({\it top panel} in Fig. \ref{fig:Slopes}), 
and $J$ versus $V-J$ ({\it bottom panel}) CMDs
in which the location of the RC magnitude and color {\it peaks} are shown.
Error bars include uncertainties in the RC peak and photometric calibration.
The least-squares fits to the data points
provide us with the slope in the CMDs,
$A_{V}/E_{V-J} = 1.255 \pm 0.004$ and 
$A_{J}/E_{V-J} = 0.225 \pm 0.005$.

%%%%---------------------------------
\begin{figure}[t]
  \begin{center}
   \epsscale{.70}
   \plotone{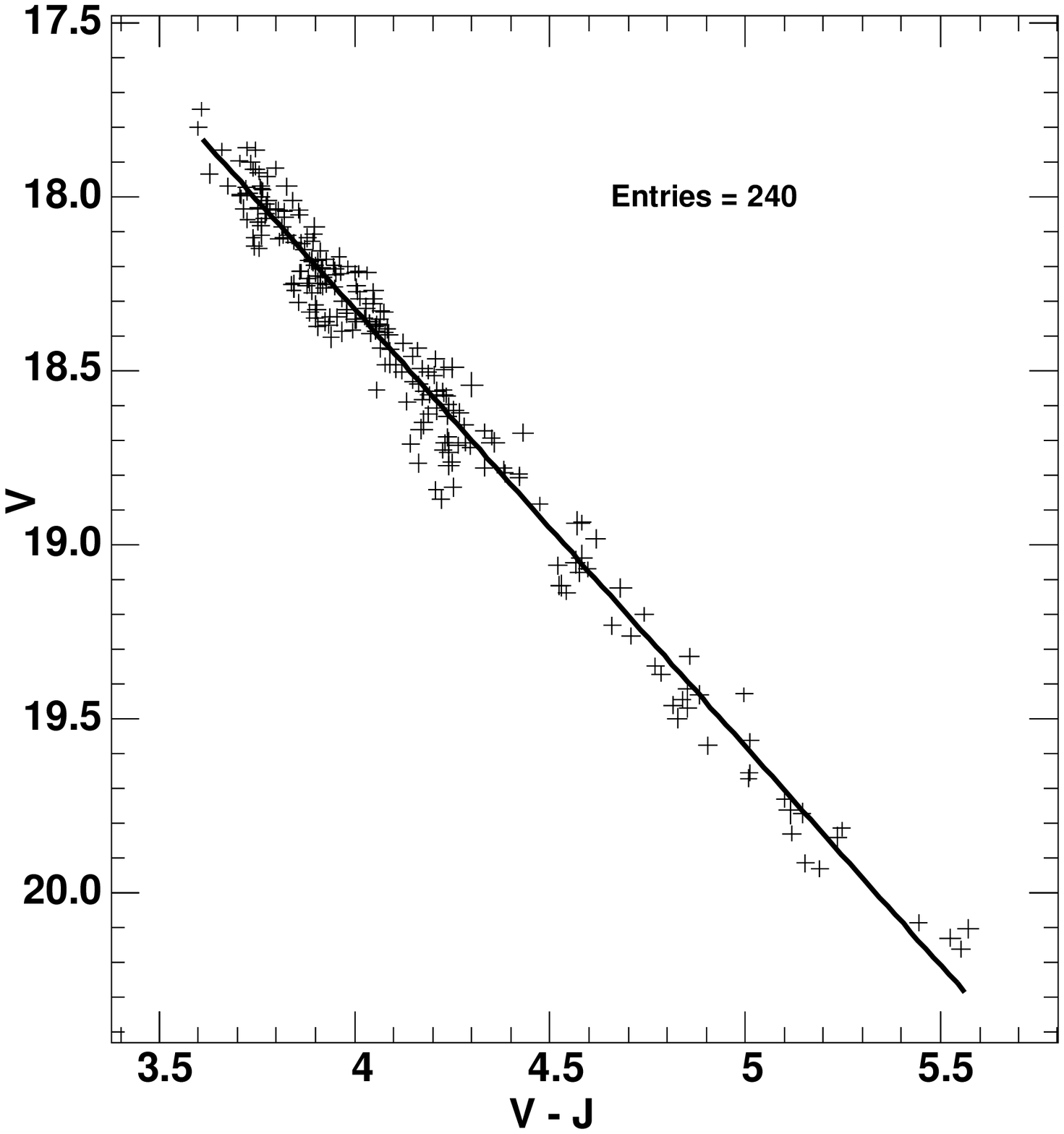}
   \epsscale{.70}
   \plotone{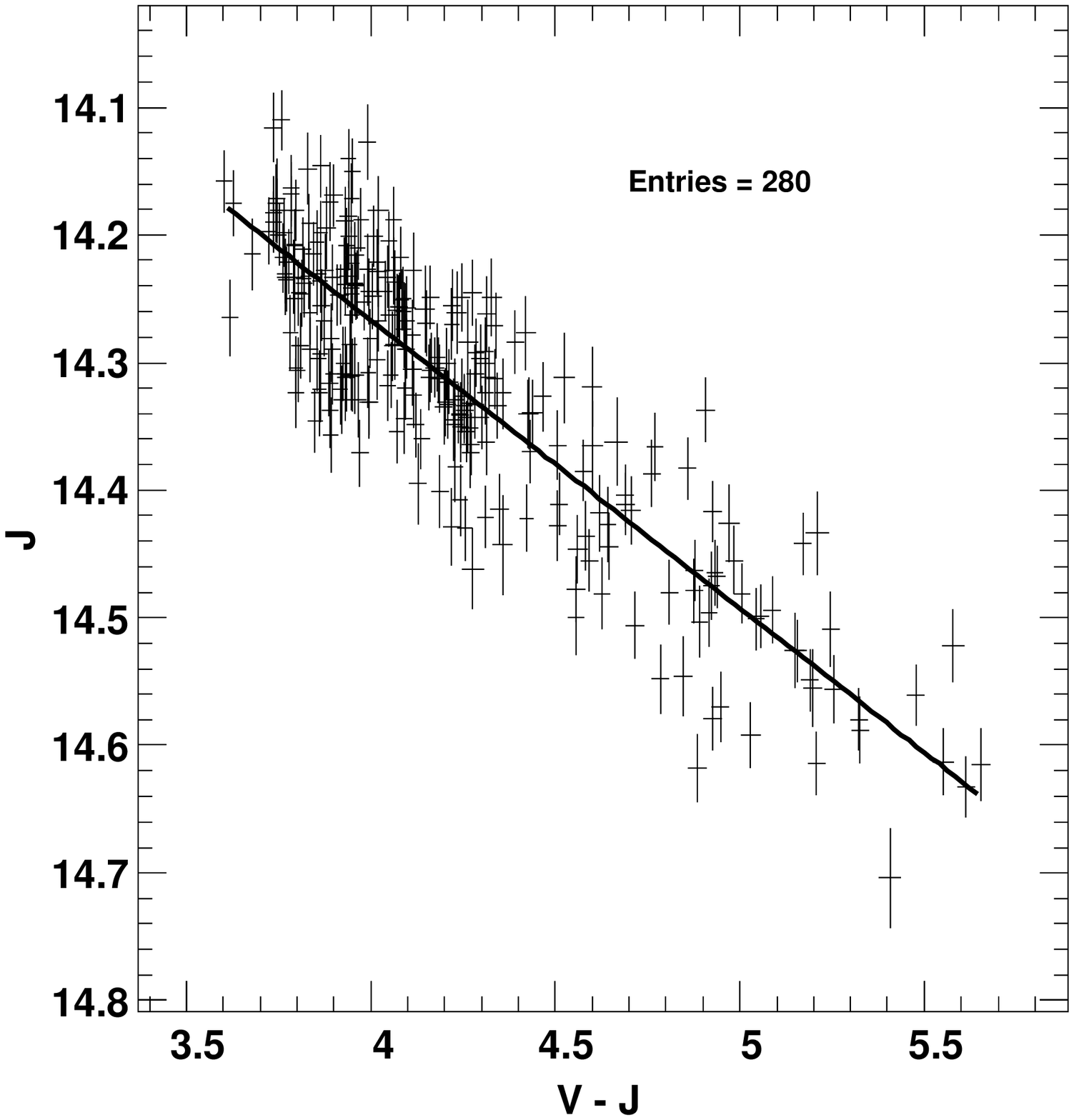}
   \end{center}
    \caption {
	Location of RC peaks in $V$ versus $V-J$ ({\it top})
	and  $J$ versus $V-J$ ({\it bottom}) CMDs.
	The solid lines are the least-squares fits to the data points.
   }
   \label{fig:Slopes}
\end{figure}

The errors in the slopes obtained by using the least-squares fit 
seem to be underestimated 
because the dispersion of the data points is large compared with the error bars,
which is particularly noticeable in the $J$ versus $V-J$ CMD.
Hence we estimated the errors of the slopes
by fixing the $\chi^2$ of the fit to equal the number of degrees of freedom,
$\chi^2 / \mathrm{dof} = 1$,
under the assumption that the errors are all equal.
Application of this procedure to the data in both CMDs
results in
$A_{V}/E_{V-J} = 1.251 \pm 0.014$ and 
$A_{J}/E_{V-J} = 0.225 \pm 0.007$.

These ratios of total to selective extinction provide 
the ratio of total extinction $A_{J}/A_{V}$, 
with a simple algebra of 
$((A_{V}/E_{V-J})-1) / (A_{V}/E_{V-J})$ and
$(A_{J}/E_{V-J}) / ((A_{J}/E_{V-J})+1)$,
yielding {$A_{J}/A_{V} = 0.201 \pm 0.011$ (from $A_{V}/E_{V-J}$) 
and $A_{J}/A_{V} = 0.184 \pm 0.006$ (from $A_{J}/E_{V-J}$).
Therefore we obtain the weighted mean and error of them,
$A_{J}/A_{V} = 0.188 \pm 0.005$.
However, the difference of the two values is 0.017,
and the combined sigma is $\sqrt{0.011^2+0.006^2}=0.013$.
Hence $A_{J}/A_{V}$ could have an error of an order of 0.01.

\section{Discussion}
\label{sec:Disc}

\subsection{Reliability of Our Data Sets}

The faintest data points in the CMDs (Fig. \ref{fig:Slopes}) 
are $V=20.2$ and $J=14.7$.
\citet{Udal03} set a safety margin and regarded his points 
as complete to the limit of $V=20.7$, 
which is still fainter than our faintest data point by 0.5 mag.  
As described in section \ref{sec:Obs},
the limiting magnitude in the $J$ band is 16.8,
also much fainter than $J=14.7$.
Therefore these margins seem to be large enough 
to have the mean magnitude of RC stars reliably measured.

The peak magnitude of RC stars could be altered by the dropping completeness.
Therefore we have checked the completeness and the change of the peak magnitudes.
First, we have made experiments in which we added artificial stars
of various known magnitudes to our original images,
and subjected them to the same procedure described in \S 2.1.
The detection rates drop as the magnitude becomes fainter;
$\approx 95 \%$ at $J = 14$,
$\approx 85 \%$ at $J = 15$,
$\approx 70 \%$ at $J = 16$.
Second, we have reconstructed luminosity functions 
with the detection rates compensated,
and fitted the RC peaks again.
Comparing the RC peak magnitudes with reconstructed ones,
we have obtained that the mean difference between them
is 0.017 mag, and its rms is 0.006.
The rms is very small compared to the error of each data points in the CMDs,
typically 0.02-0.03 mag.
We cannot find clear dependence of the difference on the peak magnitude,
and on the number of stars.
Hence we can conclude that 
the completeness effect does not change the slopes in the CMDs.
%Since the peak magnitudes depend on the extinction,
%and the crowding of stars also denends on the extinction,

The zero-point uncertainty should be checked
to derive the slope in CMDs reliably.
We examined the internal consistency
of the duplicate sources in overlapping regions of adjacent fields.
Histograms of mean magnitude difference with the sources 
of photometric error less than 0.05 mag for each field set
are shown by the left panel in Fig. \ref{fig:DifNext_2M}.
The differences of adjacent fields in the direction of R.A. and Dec.
are shown by white and hatched histograms, respectively.
We determined the rms of the magnitude difference to be less than 0.02 mag.
For another check, we have made comparisons with the 2MASS catalog
in the $J$ band.
The histogram of the mean difference between the 2MASS and IRSF $J$ magnitudes
for each field
is shown by the right panel in Fig. \ref{fig:DifNext_2M}.
The mean and rms variance of the histogram are 0.009 mag and 0.015 mag, respectively.
The rms variance is similar to the zero-point uncertainty we derived from 
the observation of the standard stars.
We therefore  conclude that
the systematic error in our data sets is about 0.02 mag.  
Note that, in this study, 
a systematic error does not come from {\it absolute} zero-point uncertainty,
but from {\it relative} one.
Even if the absolute magnitudes of RC stars 
has a systematic offset,
the resultant plots shown in Fig. \ref{fig:Slopes}
move as a whole with the slope unchanged.
Also, the right panel in Fig. \ref{fig:DifNext_2M} shows that
the absolute magnitude offset in the $J$ band is small.

%%%%%%---------------------------------
\begin{figure}
  \begin{center}
   \epsscale{0.7}
    \plotone{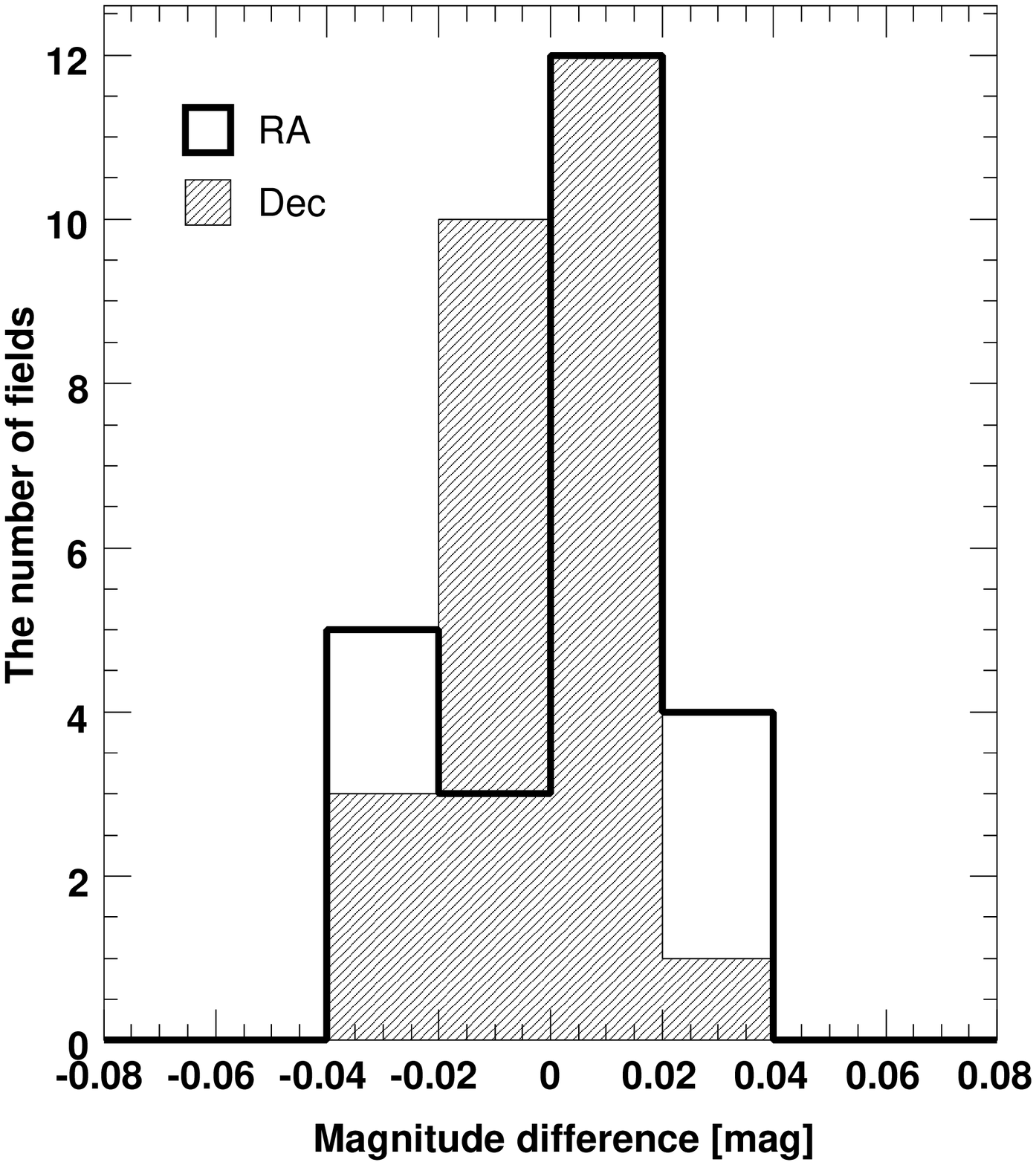}
    \epsscale{0.7}
    \plotone{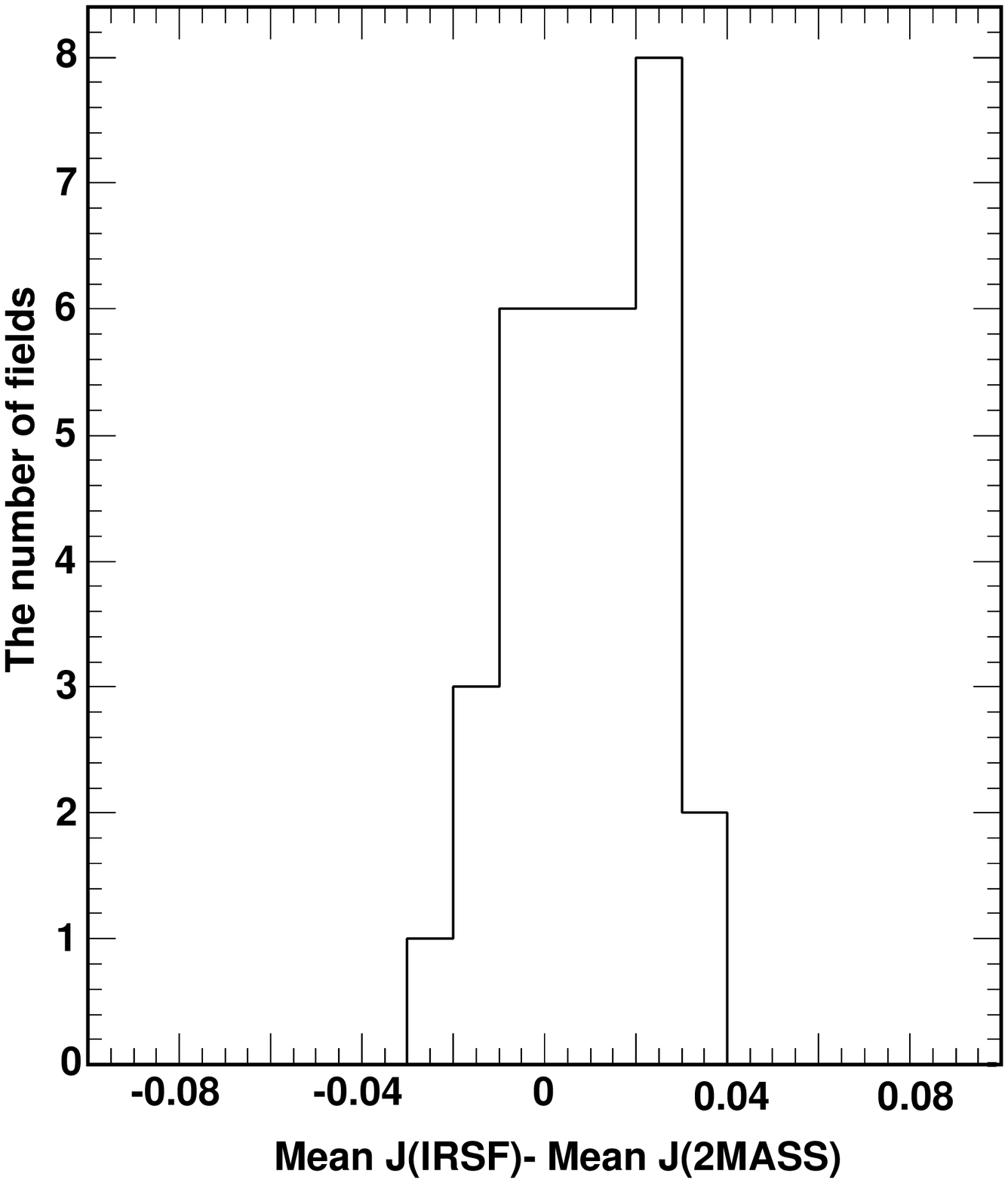}
   \end{center}
   \caption {
     ({\it Top})
	Histograms of mean magnitude differences of adjacent fields 
	with the sources of photometric error less than 0.05 mag.
	The differences in the direction of R.A. and Dec.
	are shown by {\it white} and {\it hatched} histograms, respectively.
	The mean difference and rms of the histograms are
	-0.002 and 0.016 (R.A.), and 0.002 and 0.019 (Dec).
     ({\it Bottom})
	Histogram of mean magnitude difference between 
	2MASS and IRSF sources in the $J$ band.
	The mean and rms are 0.009 and 0.015, respectively.
   }
  \label{fig:DifNext_2M}
\end{figure}

\subsection{Error Estimates of $R_\lambda$}

The difference of distance to RC stars 
in the observed area could be
a factor of the systematic error,
which become large
in the situation that the distance is correlated
with interstellar extinction.
However, the distance changes along the Galactic longitude
due to the presence of the bar structure \citep[e.g.,][]{Nakada91},
and the extinction seems to change along the Galactic latitude 
in our observed area \citep{Sumi04},
suggesting uncorrelation between them.
The extent of the observed area in the Galactic longitude is only 1\fdg1,
which leads to an only $\sim 0.03$ mag difference in
the brightness of RC stars \citep{Nishi05},
also suggesting the small systematic error by the difference of the distance.

Although population effect of the RC stars is
a source of the systematic error,
the dependence of the RC brightness on the metallicity is very weak,
and the metallicity gradient is expected to be very small
in our small observed area \citep{Udal02,Nishi05}.
In addition, the metallicity should be again correlated
with the interstellar extinction, 
and this situation seems to be unlikely.

To examine the systematic errors for the population effect and 
the different distance of RC stars,
we made linear fits to the data points in the $J$ versus $V-J$ CMD
for four OGLE fields separately,
keeping the same slope (0.225) but changing the intersect.
The intersects we obtained are 13.37, 13.37, 13.35, and 13.38
for SC37, 3, SC4, and SC39, respectively.
Next, to check the dependence of the intersect's deviation
on the slope value, we also made linear fits for the four fields,
with changing the slope values between 0.211 ($0.255-2\sigma$) and 0.239 ($0.255+2\sigma$).
We could not find any systematic trend, 
and they are within the error of the intersect, 
$13.37 \pm 0.03$ for all subfields.
We thus conclude that
the difference of the distance and population effect
do not affect our results.

The change of the effective wavelength in the $J$ band is very small.
As shown in the bottom panel of Fig. \ref{fig:Slopes},  
the peak magnitudes of RC stars are distributed 
between $m_J = 14.1$ and $14.7$.
Using the distance modulus of 14.38 toward the GC \citep{Nishiyama06b}
and the absolute magnitude $M_J \approx -0.3$ of RC stars \citep{Bonatto04}, 
$A_J$ can be estimated to be $0.02 \lesssim A_J \lesssim 0.68$.
In this range of extinction, 
the effective wavelength in the $J$ band for typical bulge RC stars
changes by only 0.002$\mu$m \citep[see][\S 5.1, for more details]{Nishiyama06b},
and the change of effective wavelength is thus negligible.

\subsection{Extinction $A_{\lambda}/A_V$ toward Galactic Bulge}

In the wavelength range $0.5 - 0.9 \mu$m, 
extinction toward the Galactic bulge has been characterized by 
a ``steep'' curve whose ratio of total to selective extinction is small.
The steep extinction curve was introduced to explain 
the anomalous $(V-I)_0$ color of RC stars in Baade's window
\citep[e.g.,][]{Popo00,Gould01}.
The lower value of the ratio of total to selective extinction 
was also reported 
from the MACHO $V$ and $R$ photometry 
\citep[$A_V / E_{V-R} \sim 3.5$,][]{Popo03},
and from the OGLE $V$ and $I$ photometry 
\citep[$R_{VI} = A_V / E_{V-I} \sim 2.0$,][]{Udal03, Sumi04}
toward the Galactic Bulge.   
The empirical analytic formula of CCM or \citet{Fitz99} corresponding to 
the extinction toward the Galactic bulge has 
the single parameter $R_{V}$ of $\approx 2$, 
which is much lower than the average value 3.1
for diffuse regions in the local interstellar medium.  

The CCM formula uses a simple power law $\lambda^{-1.61}$ 
in the wavelength region 
$> 0.9 \mu$m and it seems independent of $R_V$.  
However, since $A_V$ depends on $R_V$, 
the ratios $A_V / E_{V-J}$ and $A_J / E_{V-J}$ 
are dependent on $R_V$, and these ratios 
derived in our work\footnote{ 
$R_{V} \sim 1.8$ was obtained to reproduce our result
$A_{J}/A_{V} = 0.193$
by using the CCM formula.
}
correspond to 
$R_{V} \sim 1.8$, which is also very small.  
The reddening vectors in the $V$ versus $V-J$ CMD
for the case of $R_V=3.1$ and our results are plotted 
in Fig. \ref{fig:CMD}.  
The figure shows a clear difference in the reddening direction.  

Small values of $R_{V}$ are generally considered to 
indicate the prevalence of small dust grains which affect 
the extinction curve in the ultraviolet - optical wavelengths.  
Although a substantial number of lines of sight 
with low $R_V$ values are found especially 
at high Galactic latitude \citep[e.g.,][]{Larson96,Larson05}, 
only a few lines of sight exist with $R_V<2.0$ \citep{Szomoru99,Larson05}.  

Next, we try to extend the extinction curve to $2 \mu$m.  
Paper I determined the dependence of the interstellar extinction
in the $J$, $H$, and $K_S$ bands toward the GC,
and thus we can combine it with the result obtained in this paper
to determine $A_{\lambda}/A_V$ in the $H$, and $K_S$ bands.  
Here we should take into account the variation of the extinction law
in different lines of sight,
because the area observed in paper I
($\mid l \mid \la 2\fdg0$ and $0\fdg5 \la \mid b \mid \la 1\fdg0$)
does not overlap with that of this study.
The variation of $A_{K_S}/E_{H-K_S}$ was estimated to be as large as $\sim 7 \%$
in the region of $4\degr \times 2\degr$ at the GC (paper I),
and hence we adopt this value as a variation of $A_{J}/A_{V}$,
resulting $A_{J}/A_{V} = 0.188 \pm 0.014$ 
where $0.014 = \sqrt{ (0.005)^2 + (0.188 \times 0.07)^2}$.
By using the ratio $ A_{J} : A_{H} : A_{K_S} = 1 : 0.573 \pm 0.009 : 0.331 \pm 0.004 $ (paper I),
we obtain
$ A_{J}/A_{V} : A_{H}/A_{V} : A_{K_S}/A_{V} = 0.188 \pm 0.014 : 0.108 \pm 0.008 : 0.062 \pm 0.005$.

\begin{table}[h]
 \begin{center}
  \caption{The wavelength dependence of the interstellar extinction.}
 \vspace{0.5cm}
  \begin{tabular}[c]{c|lll}\hline \hline
   \dorule \uprule & IRSF
   & vdH\tablenotemark{(1)} & CCM\tablenotemark{(2)} \\ \hline
   $A_{V}/E_{V-J}$ & $1.251\pm0.014$ & 1.325  & 1.393  \\
   $A_{J}/E_{V-J}$ & $0.225\pm0.007$ & 0.325 & 0.393 \\ \hline
   $A_J/A_V$ & $ 0.188\pm0.014\tablenotemark{(3)} $ & 0.245 & 0.282 \\
   $A_H/A_V$ & $ 0.108\pm0.008\tablenotemark{(3)} $ & 0.142 & 0.190 \\
   $A_{K_S}/A_V$ & $ 0.062\pm0.005\tablenotemark{(3)} $ & 0.088 &  0.118  \\
  \hline
  \end{tabular}
  \label{tab:ratios}
  \tablerefs{
  (1) \citet{vdH49}; \\%%(2) \citet{RL85};
  (2) \citet{CCM89} for the case of $R_V=3.1$. \\
  (3) The error coming from variation of extinction law in\\ different lines of sight is included. }
 \end{center}
\end{table}

The resultant $A_{\lambda}/A_V$ in the $J H K_S$ wavelength range is a 
steeply declining function.  
As shown in Table \ref{tab:ratios}, the CCM curve (for $R_V =3.1$), 
which is based on \citet{RL85}, decreases much more slowly toward the 
longer wavelengths.  
In particular, the $K_S$ band extinction is slightly greater than 
one tenth of the visual extinction $A_V$ in the CCM curve, 
which contrasts with the \citet{vdH49} curve, where $A_{K_S}$ is 
slightly less than one tenth of $A_V$.  
The derived extinction toward the Galactic bulge decreases more steeply 
as the wavelength increases.  
The steep decrease is rather striking, but 
it was already evident in \citet{Messineo05} and paper I, 
where a steep extinction power law index $\alpha \approx 2.0$,
consistent with 
the polarization law up to $\sim 2.5 \mu$m \citep{Nagata94},
was proposed for the GC extinction.  
To confirm this, 
deep optical and infrared observations in the $V$ to $K_S$ bands
for the same fields with appropriate extinction would be important.  

\acknowledgements

We thank the staff at the South African Astronomical Observatory (SAAO)
for their support during our observations.
The IRSF/SIRIUS project was initiated and supported by Nagoya
University, the National Astronomical Observatory of Japan
in collaboration with the SAAO.
This work was supported by KAKENHI,
Grant-in-Aid for Young Scientists (B) 19740111,
and in part by the Grants-in-Aid for the 21st Century 
COE ``The Origin of the Universe and Matter: 
Physical Elucidation of the Cosmic History'' from the MEXT of Japan.
This publication makes use of data products from the Two Micron All Sky Survey, 
which is a joint project of the University of Massachusetts and 
the Infrared Processing and Aeronautics and 
Space Administration and the National Science Foundation.

\label{sec:Conc}

%--------------------------------------------------------

\end{document}